\documentclass[aip,jcp, reprint,groupedaddress]{revtex4-2}%
\usepackage{graphicx}
\usepackage{tabularx}
\usepackage{dcolumn}
\usepackage{longtable}
\usepackage{tensor}
\usepackage{color}
\usepackage{placeins}
\usepackage{bm}
\usepackage{amsmath}
\usepackage{amsfonts}
\usepackage{amssymb}
\usepackage{float}
\usepackage[urlcolor=white]{hyperref}%
\providecommand{\U}[1]{\protect\rule{.1in}{.1in}}
\providecommand{\U}[1]{\protect\rule{.1in}{.1in}}
\graphicspath{{./Figures/}}
\begin{document}
\title{Resonance Raman spectroscopy from ab initio Hagedorn wavepacket dynamics}
\author{Davide Barbiero}
\thanks{These authors contributed equally to this work.}
\author{L\'{e}a Zupan}
\thanks{These authors contributed equally to this work.}
\author{Ji\v{r}\'i J. L. Van\'i\v{c}ek}
\email{jiri.vanicek@epfl.ch}
\affiliation{Laboratory of Theoretical Physical Chemistry, Institut des Sciences et
Ing\'enierie Chimiques, Ecole Polytechnique F\'ed\'erale de Lausanne (EPFL),
CH-1015, Lausanne, Switzerland}
\date{\today}

\begin{abstract}
We present a practical, ab initio time-dependent method using Hagedorn
wavepackets to simulate resonance Raman (RR) spectra of polyatomic molecules.
Hagedorn functions---Gaussians multiplied by specific polynomials---are used
to represent RR initial and final states because these functions are exact
solutions to the time-dependent Schr\"{o}dinger equation for at-most-quadratic
potentials and can be propagated at zero cost beyond that of propagating the
guiding Gaussian. Using efficient recursive formulae to compute overlaps
between Hagedorn wavepackets, we can evaluate RR excitation profiles for
arbitrary spectral signals, such as fundamental, overtone, combination, and
hot bands. We then construct the Stokes and anti-Stokes RR spectra from these
profiles. We first validate the method in a two-dimensional displaced,
distorted, and Duschinsky-rotated harmonic model against numerically exact
split-operator calculations. Then, we apply the method to compute RR spectra
of anthracene by performing dynamics on a 66-dimensional harmonic potential
energy surface constructed from density functional theory calculations.

\end{abstract}
\maketitle

\footnotetext{These authors contributed equally}



In resonance Raman (RR) scattering, an incident photon of frequency
$\omega_{I}$ excites the initial vibrational wavepacket $\psi_{i}$ from the
ground, $g$, to an excited electronic state, $e$. Subsequent scattering of a
photon of frequency $\omega_{S}$ returns the system to the ground electronic
state and vibrational state $\psi_{f}$%
.~\cite{Albrecht:1961,book_Tannor:2007,book_Heller:2018} This technique has
become a powerful tool in chemistry,\cite{Gustin_Franco:2023,Malenfant_Frenette:2024,Philipp_Begusic:2026} materials
science,\cite{Wu_Tan:2018} and biology,\cite{Dodo_Mikiko:2022} due to two
major advantages: First, RR spectroscopy is much more sensitive to
ground-state vibrational structure than non-resonant Raman spectroscopy
because the resonance condition enhances the scattered intensity by several
orders of magnitude;~\cite{Holtum_Schlucker:2021} second, RR spectroscopy
yields richer information about excited-state structure and dynamics than
linear absorption spectroscopy because the scattered intensity depends on both
the ground- and excited-state potential energy
surfaces.~\cite{Inagaki_Miyazawa:1974}

The key quantity in RR spectroscopy is the polarizability tensor between the
initial and final vibrational states. Computational strategies for evaluating
this tensor can be divided into two main classes: In time-independent
methods,~\cite{Peticolas_Rush:1995,Guthmuller_Gonzalez:2010,Heller_Kaxiras:2016} the
polarizability tensor is expressed in terms of Franck--Condon integrals, as
introduced by Kramers, Heisenberg, and
Dirac.~\cite{Kramers_Heisenberg:1925,Dirac:1927} Recursive schemes have been
developed to evaluate these integrals,~\cite{Sharp_Rosenstock:1964} together
with prescreening techniques~\cite{Santoro_Barone:2011,Egidi_Barone:2014} that
identify dominant transitions and reduce the number of vibrational overlaps
that must be evaluated. Instead, in time-dependent
methods~\cite{Ben-Nun_Martinez:1999_v2,Ovchinnikov_Apkarian:2001,Baiardi_Barone:2014_v2,Bessone_Spezia:2023}
the Kramers--Heisenberg--Dirac equation is reformulated in terms of nuclear
wavepacket dynamics, as introduced by Heller and
coworkers.~\cite{Lee_Heller:1979,Heller_Tannor:1982,Tannor_Heller:1982} Due to
its simplicity and efficiency, the time-dependent formulation of RR scattering
has proved valuable both computationally and
conceptually.~\cite{Tannor:1988,Mattiat_Luber:2021}

Among the many semiclassical methods that can be used to propagate nuclear
wavepackets,~\cite{Miller:2001,book_Heller:2018,Vanicek:2023,Conte_Ceotto:2024} we turn to Hagedorn wavepacket
dynamics.~\cite{Hagedorn:1998,Lasser_Lubich:2020} This approach was introduced
only recently in vibronic spectroscopy for the computation of single vibronic
level fluorescence spectra.~\cite{Zhang_Vanicek:2024a,Zhang_Vanicek:2025}
Here, we describe how Hagedorn wavepackets can be used to compute RR spectra.
We first show that this method is exact in many-dimensional displaced,
distorted, and Duschinsky-rotated harmonic models by comparing it with
numerically exact quantum split-operator calculations. Then, we combine
Hagedorn wavepackets with ab initio time-dependent density functional theory
to compute RR signals of anthracene.



Within the electric dipole approximation, Condon approximation, and
second-order time-dependent perturbation theory, the RR scattering
cross-section, integrated over all directions and polarizations, can be
computed as the half Fourier
transform~\cite{Lee_Heller:1979,book_Tannor:2007}
\begin{equation}
\sigma_{fi}(\omega_{I})=\frac{8i\pi\omega_{I}\omega_{S}^{3}}{9\hbar c^{4}}%
\mu^{2}\int_{0}^{\infty} C_{fi}(t) e^{i(\omega_{I}+E_{g,i}/\hbar)t}dt
\label{eq:cross-sec}%
\end{equation}
of the nuclear wavepacket cross-correlation function
\begin{equation}
C_{fi}(t)=\langle\psi_{f}|e^{-i\hat{H}_{e}t/\hbar}|\psi_{i}\rangle,
\label{eq:corr}%
\end{equation}
where the wavepacket $\psi_{i}$ evolves under the time-dependent
Schr\"{o}dinger equation (TDSE) with the final-state vibrational Hamiltonian
$\hat{H}_{e}=T(\hat{p})+V_{e}(\hat{q})$. The kinetic term $T(p)=p^{T} \cdot
m^{-1} \cdot p/2$ depends on the momentum $p$ and the real symmetric mass
matrix $m$, and the Born-Oppenheimer potential energy surface $V_{e}(q)$
depends only on position $q$. In Eq.~(\ref{eq:cross-sec}), $\mu=\lVert\vec
{\mu}_{eg}\rVert$ is the magnitude of the electronic transition dipole moment
evaluated at the ground-state equilibrium geometry and $E_{g,i}$ is the
vibronic energy of state $\psi_{i}$ before photon absorption.

In a harmonic ground-state surface
\begin{equation}
V_{g}(q) := q^{T} \cdot\kappa_{g} \cdot q/2, \label{eq:V_g}%
\end{equation}
with the force constant $\kappa_{g}:=\text{Hess}\,V_{g}(q_{0})$ at the
equilibrium geometry $q_{0}=0$, the initial, $\psi_{i}$, and final, $\psi_{f}%
$, vibrational wavefunctions with vibrational quantum numbers $K=(K_{1}%
,\dots,K_{D})\in\mathbb{N}_{0}^{D}$ can be represented as Hagedorn
functions~\cite{Hagedorn:1998,Lasser_Lubich:2020}
\begin{equation}
\varphi_{K} = (K!)^{-1/2}(A^{\dag})^{K}\varphi_{0}. \label{eq:phi_k}%
\end{equation}
These functions are constructed from a $D$-dimensional, normalized,
complex-valued Gaussian wavepacket
\begin{align}
\varphi_{0}[\Lambda_{t}](q) &= \frac{1}{(\pi\hbar)^{D/4}\sqrt{\text{det}%
(Q_{t})}} \nonumber\\
&~~~\times\text{exp} \left[  \frac{i}{\hbar} \left(  \frac{1}{2}x^{T} \cdot
P_{t} \cdot Q_{t}^{-1} \cdot x + p_{t}^{T} \cdot x + S_{t}\right)  \right]
\label{eq:phi_0}%
\end{align}
(representing the ground vibrational wavefunction) by applying the Hagedorn
raising operator~\cite{Hagedorn:1998,Lasser_Lubich:2020}
\begin{equation}
A^{\dag} := \frac{i}{\sqrt{2 \hbar}} \left[  P_{t}^{\dag} \cdot(\hat{q} -
q_{t}) - Q_{t}^{\dag} \cdot(\hat{p} - p_{t})\right]  . \label{eq:Adag}%
\end{equation}
In Eq.~(\ref{eq:phi_k}), we used the multi-index notation for $K!:=K_{1}!\dots
K_{D}!$ and\newline$(A^{\dag})^{K}:=(A_{1}^{\dag})^{K_{1}}\dots(A_{D}^{\dag
})^{K_{D}}$, where $A_{j}^{\dag}$ is the $j$-th component of the raising
vector operator $A^{\dag}$. In Eq.~(\ref{eq:phi_0}), $x:=q-q_{t}$ is the
shifted position and $\Lambda_{t}=(q_{t},p_{t},Q_{t},P_{t},S_{t})$ is a set of
time-dependent parameters, where $q_{t}$ and $p_{t}$ represent the position
and momentum of the center of the Gaussian wavepacket~(\ref{eq:phi_0}),
$Q_{t}$ and $P_{t}$ are two complex $D$-dimensional matrices that factorize
the width matrix $C_{t}=P_{t}\cdot Q_{t}^{-1}$ of the Gaussian,~\cite{Heller:1976a,Hagedorn:1980} and $S_{t}$ is a real phase related to the classical action.

Hagedorn functions $\varphi_{K}[\Lambda_{t}]$, just like thawed
Gaussians~\cite{Heller:1975,Vanicek:2023} $\varphi_{0}[\Lambda_{t}]$, are
exact solutions to the TDSE in a displaced, distorted, and Duschinsky-rotated
harmonic potential~\cite{Lasser_Lubich:2020,Vanicek_Zhang:2025_v2}
\begin{equation}
V_{e}(q):=V_{\text{eq}}+(q-q_{\text{eq}})^{T}\cdot\kappa_{e}\cdot
(q-q_{\text{eq}})/2,\label{eq:V_e}%
\end{equation}
The parameters $\Lambda_{t}$ of the Gaussian associated with the Hagedorn
function $\varphi_{K}[\Lambda_{t}]$ evolve according to the equations
\begin{align}
\dot{q}_{t} &  =m^{-1}\cdot p_{t},\label{eq:qEOM}\\
\dot{p}_{t} &  =-V_{e}^{\prime}(q_{t})=-\kappa_{e}\cdot(q_{t}-q_{\text{eq}%
}),\label{eq:pEOM}\\
\dot{Q}_{t} &  =m^{-1}\cdot P_{t},\label{eq:QEOM}\\
\dot{P}_{t} &  =V_{e}^{\prime\prime}(q_{t})\cdot Q_{t}=-\kappa_{e}\cdot
Q_{t},\label{eq:PEOM}\\
\dot{S}_{t} &  =T(p_{t})-V_{e}(q_{t}),\label{eq:SEOM}%
\end{align}
while---remarkably---the multi-index $K$ does not change during
propagation.~\cite{Lasser_Lubich:2020,Vanicek_Zhang:2025_v2}

The cross-correlation function~(\ref{eq:corr}), needed for evaluating the RR
profile, requires evaluating the overlap
\begin{align}
M_{JK^{\prime}}  &  :=\langle\varphi_{J}(\Lambda_{0})|\varphi_{K}(\Lambda
_{t})\rangle\nonumber\\
&  =\langle\psi_{f}(0)|\psi_{i}(t)\rangle\nonumber\\
&  =C_{fi}(t)
\end{align}
of Hagedorn functions with different Gaussian centers. This scalar product can
be evaluated efficiently in terms of the simple overlap $M_{00^{\prime}}$ of
Gaussians with different parameters using the algebraic recursive expressions
from Ref.~\onlinecite{Vanicek_Zhang:2025_v2}. Remarkably, Hagedorn wavepackets
enable the evaluation of RR excitation profiles from any initial vibrational
level using a single trajectory $\Lambda_{t}$ of the common guiding Gaussian.

In the Supporting Information, we describe how the general recursive
expressions simplify substantially if one is interested in low excitations
(i.e., fundamentals, first overtones, and first combination bands) in the
zero-temperature RR spectrum.



\begin{figure}
\includegraphics{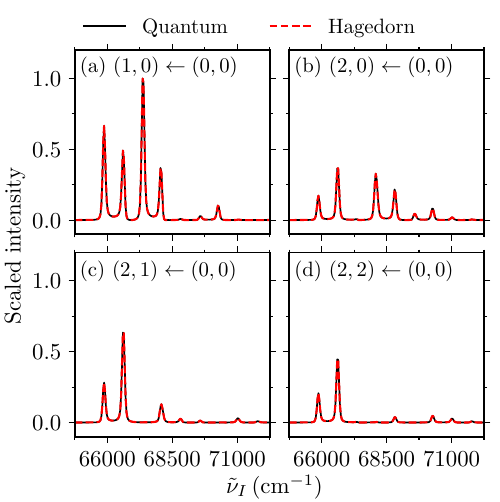}
\caption{Comparison of the resonance Raman excitation profiles computed using quantum split-operator algorithm and Hagedorn wavepacket dynamics in a displaced, distorted, and Duschinsky-rotated two-dimensional harmonic system. Profiles for fundamental [panel (a)], first-order overtone [panel (b)], and two combination bands [panels (c) and (d)] are shown.}
\label{fig:HO_exc_prof}
\end{figure}

We begin by considering a displaced, distorted, and Duschinsky-rotated
two-dimensional harmonic system. Even for such a simple model, the spectrum
can be surprisingly complicated due to the two-dimensional nature of RR
spectroscopy.~\cite{Tannor_Heller:1982,Tannor:1988} Nevertheless, numerically
exact quantum calculations remain readily accessible, allowing direct
verification of spectra computed using Hagedorn wavepacket dynamics.
Figure~\ref{fig:HO_exc_prof} compares the RR excitation profiles obtained with
the two methods. These profiles are constructed by monitoring the scattered
intensity at a fixed scattered wavenumber $\tilde{\nu}_{S}$ as a function of
the incident wavenumber $\tilde{\nu}_{I}$. The algebraic recursive
expressions~\cite{Vanicek_Zhang:2025_v2} for the overlap of Hagedorn functions
enable the efficient evaluation of RR excitation profiles for any final
vibrational level. As an illustration, we present profiles for final states
containing 1, 2, 3, and 4 vibrational quanta. In all cases displayed in
Fig.~\ref{fig:HO_exc_prof}, we assumed the zero-temperature approximation, in
which the initial state was the vibrational ground state, i.e., a Gaussian,
and only the final state was general Hagedorn function. The spectra obtained
from Hagedorn dynamics (which here reduces to Gaussian wavepacket dynamics
except for the overlap with the final state) are visually indistinguishable
from those computed using the grid-based quantum approach. Moreover, absolute
differences are extremely small ($<10^{-11}$, not shown).

\begin{figure}
\includegraphics{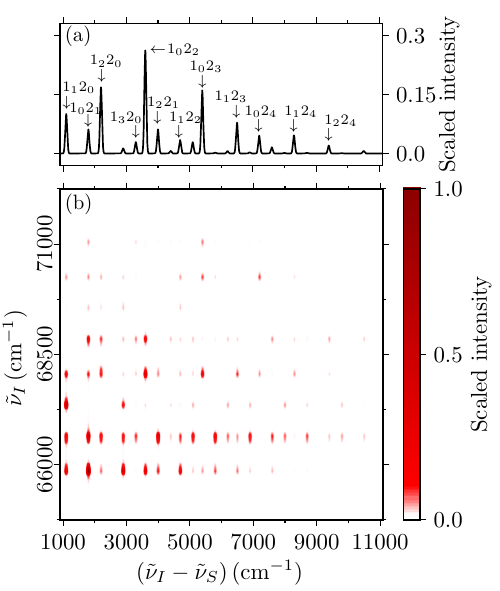}
\caption{Resonance Raman spectra computed using Hagedorn wavepacket dynamics in a displaced, distorted, and Duschinsky-rotated two-dimensional harmonic system. (a) Resonance Raman spectrum for $\tilde{\nu}_{I}=68120\,\text{cm}^{-1}$. (b) Two-dimensional resonance Raman spectrum.}
\label{fig:HO_2D_spec}
\end{figure}

In contrast to excitation profiles, Raman spectra are obtained by fixing
$\tilde{\nu}_{I}$ and looking at the scattered intensity as a function of the
Raman shift $\tilde{\nu}_{S}-\tilde{\nu}_{I}$. Panel (a) of
Fig.~\ref{fig:HO_2D_spec} shows one such Raman spectrum. The three most
intense peaks correspond to overtones. By carefully selecting the incident
frequency, different transitions, e.g., fundamental or combination bands, can
be selectively enhanced in the spectrum.~\cite{Tannor:1988} This effect
becomes particularly evident in panel (b), which displays the full
two-dimensional Raman spectrum at zero temperature.

\begin{figure}
\includegraphics{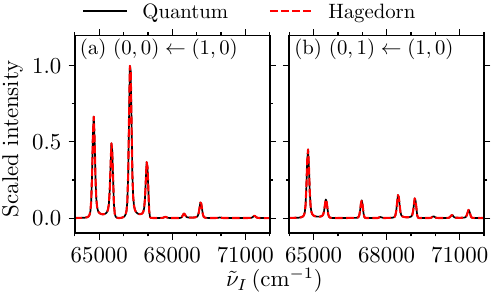}
\caption{Comparison of the resonance Raman excitation profiles computed using quantum split-operator algorithm and Hagedorn wavepacket dynamics in a displaced, distorted, and Duschinsky-rotated two-dimensional harmonic system. Profiles for anti-Stokes [panel (a)] and Stokes [panel (b)] hot bands are shown.}
\label{fig:HO_hot_bands}
\end{figure}

Whereas the calculations of zero-temperature RR spectra only require
propagating a Gaussian wavepacket, the evaluation of both Stokes and
anti-Stokes hot bands at nonzero temperatures requires propagating general
Hagedorn functions. Remarkably, the propagation of arbitrary Hagedorn
functions, which is needed for RR spectra at nonzero temperatures, can be done
using the same Gaussian guiding trajectory already employed at zero
temperature.~\cite{Zhang_Vanicek:2024a,Zhang_Vanicek:2025a} For vibrationally excited initial
states, the dynamics of Hagedorn wavepackets differs from that of simple
products of one-dimensional Hermite
functions,~\cite{Zhang_Vanicek:2024a,Zhang_Vanicek:2025a} but the
time-dependent Hagedorn basis is such that the excitation number $K$ remains
constant for each Hagedorn function, even though the function may evolve in a
complicated way. Figure~\ref{fig:HO_hot_bands} confirms that Hagedorn
wavepacket dynamics in many-dimensional harmonic systems remains
exact.\cite{Hagedorn:1998,Lasser_Lubich:2020}

Following validation, we applied this method to compute RR spectra of
anthracene by performing wavepacket dynamics on a 66-dimensional harmonic
potential energy surface derived from density functional theory calculations.
Anthracene was selected due to the availability of experimental
data~\cite{Efremov_Gooijer:2006} and previous computational
studies~\cite{DeSouza_Izsak:2019,Tapavicza:2019} confirming the adequacy of the harmonic
approximation and the weak influence of Herzberg--Teller coupling.

\begin{figure}
\includegraphics{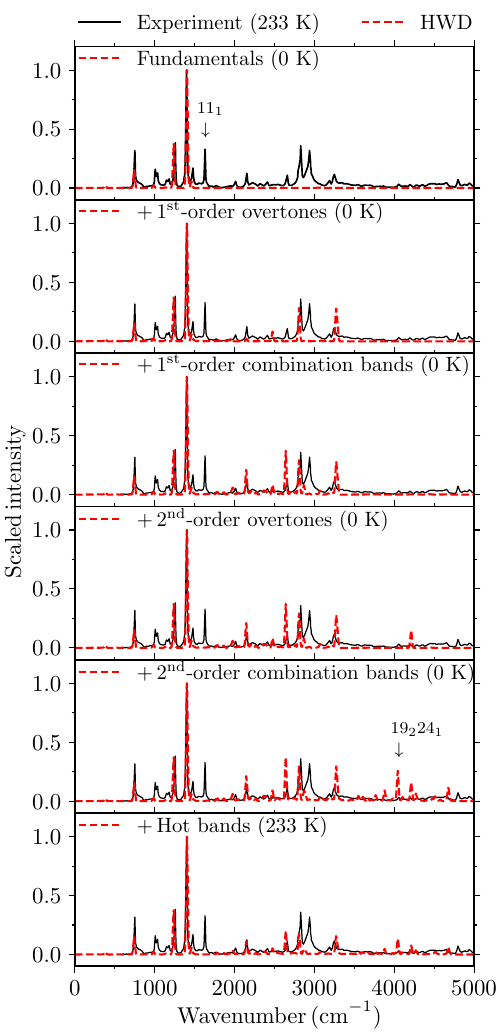}
\caption{Resonance Raman spectrum of anthracene for $\tilde{\nu}_{I}=43668\,\text{cm}^{-1}$. Spectra
computed using Hagedorn wavepacket dynamics (HWD) at 0 K with progressively increasing numbers of final
states (top five panels) and at 233 K (bottom panel) are
compared to the experimental spectrum at 233 K.~\cite{Efremov_Gooijer:2006}
The simulated spectrum at 233 K was obtained by Boltzmann-averaging the spectra of all initial vibrational states
whose excitation energies, measured relative to the zero-point energy, are lower than $2k_{B}T$.
} \label{fig:ANT_spec}
\end{figure}

Using Hagedorn wavepacket dynamics, we computed both the zero- and
finite-temperature RR spectra of anthracene while progressively increasing the
number of included final vibrational states, from fundamentals to second-order
overtones and combination bands. The resulting zero-temperature spectra are
compared with the experiment in the top five panels of Fig.~\ref{fig:ANT_spec}
for an incident wavenumber $\tilde{\nu}_{I}=43668\,\text{cm}^{-1}$. As
reported previously,~\cite{DeSouza_Izsak:2019} inclusion of higher-excited
vibrational states is crucial for accurately reproducing the experimental
spectrum. Overall, the simulated spectrum agrees well with the experimental
one and is consistent with the result obtained in
Ref.~\onlinecite{DeSouza_Izsak:2019} using a path-integral formulation. Since
both approaches are formally exact within the harmonic approximation, this
agreement is expected. In contrast to the alternative method from
Ref.~\onlinecite{DeSouza_Izsak:2019}, the Hagedorn approach enables the
efficient computation of RR spectra involving arbitrarily highly excited final
states through the recursive evaluation of cross-correlation
functions.~\cite{Vanicek_Zhang:2025_v2}

Within the zero-temperature and Condon approximations, the selection rules
provide that transitions are only allowed to vibrational levels of $a_{g}$
symmetry. Therefore, the total quantum number in non-totally symmetric modes
must be even. Together with selection rules, the harmonic model allows clear
peak assignments in the computed spectra. For example, the peak at
$4058\,\text{cm}^{-1}$ corresponds to the combination band $19_{2}24_{1}$,
while the peak at $1654\,\text{cm}^{-1}$ corresponds to the $11_{1}$
transition and is not captured in our calculation because it originates from
Herzberg--Teller coupling.

Moreover, as demonstrated for the two-dimensional harmonic oscillator model,
temperature effects can be easily included at a minor computational cost of
evaluating additional overlaps along the same trajectory already needed for
the zero-temperature calculation. In a harmonic system, hot bands occur at the
same Raman shifts as the corresponding zero-temperature transitions and
therefore only modify peak intensities. Because the excitation profiles
associated with hot bands are red-shifted relative to the zero-temperature
profiles, thermal population enhances low-frequency peaks while reducing the
relative intensity of high-frequency peaks. As shown in the bottom panel of
Fig.~\ref{fig:ANT_spec}, the dominant temperature effect in the computed
resonance Raman spectrum of anthracene is the attenuation of overtone and
combination-band intensities; inclusion of the temperature effects further
improves the agreement between the simulated and experimental spectra.

\begin{figure}
\includegraphics{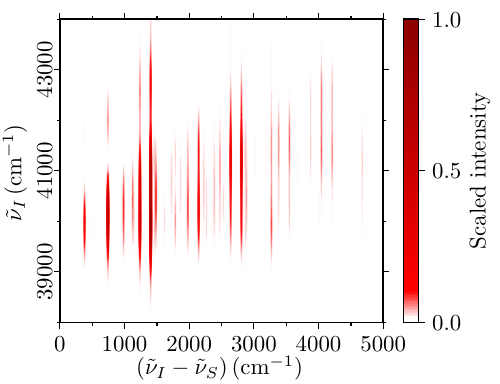}
\caption{Two-dimensional resonance Raman spectrum of anthracene computed using Hagedorn wavepacket dynamics.}
\label{fig:ANT_2D_spec}
\end{figure}

Although some experimental RR spectra of anthracene are available in the
literature,~\cite{Efremov_Gooijer:2006} to the best of our knowledge no
excitation profiles have been reported so far. Figure~\ref{fig:ANT_2D_spec}
displays the full two-dimensional Raman spectrum at zero temperature as
predicted by adiabatic harmonic Hagedorn wavepacket dynamics. As already
observed for the model system, different transitions, e.g., fundamental,
overtone, or combination bands, can be selectively enhanced in the RR spectrum
by carefully selecting the incident frequency.

In the absence of experimental excitation profiles, the broadening of the
computed profiles was estimated empirically from the experimental absorption
spectrum, exploiting the fact that in solution both processes are typically
affected by similar broadening mechanisms.~\cite{Zakaraya_Uistrup:1989} As
shown in Fig. S1 of the Supporting Information, the dominant solvent effect in
the RR spectrum of anthracene appears to be the attenuation of overtone and
combination-band intensities. In the present work, solvent effects were
treated within a polarizable continuum model; however, the efficiency of
Hagedorn wavepacket dynamics should make explicit solvent simulations
computationally feasible.


In conclusion, we have introduced a robust and efficient framework for
evaluating RR spectra using Hagedorn wavepackets. The method is exact in
global harmonic models and naturally accounts for mode-mixing and
mode-distortion effects. Notably, excitation profiles for arbitrary spectral
signals---fundamental, overtone, combination, and hot bands---are obtained via
post-processing of a single Gaussian wavepacket trajectory. Our numerical
results demonstrate that the inclusion of high-order overtones and combination
bands is essential to reproduce the spectral features of harmonic systems.
These signals become even more pronounced in anharmonic molecules;~\cite{Conte_Ceotto:2023,Begusic_Vanicek:2022} in this
context, Hagedorn wavepackets provide a flexible and systematically improvable
framework for incorporating weak
anharmonicity~\cite{Zhang_Vanicek:2025,Zhang_Vanicek:2026} because the
Hagedorn functions are exact solutions of the TDSE not only in harmonic
systems but also in general systems within the local harmonic
approximation,\cite{Hagedorn:1998,Lasser_Lubich:2020} which has proven very
useful for Gaussian wavepacket
dynamics.\cite{Heller:1975,Kletnieks_Vanicek:2023}


\section{Computational methods}

\subsection{Two-dimensional harmonic oscillator}

In our two-dimensional model system, the initial wavepackets were the
eigenstates~(\ref{eq:phi_k}) of the harmonic potential~(\ref{eq:V_g}), with
the diagonal Hessian matrix $\kappa_{g}$ corresponding to vibrational
wavenumbers $\tilde{\nu}_{1}^{\prime\prime}=1100\,\text{cm}^{-1}$ and
$\tilde{\nu}_{2}^{\prime\prime}=1800\,\text{cm}^{-1}$. These wavepackets were
excited under the sudden approximation to the displaced, distorted, and
Duschinsky-rotated surface~(\ref{eq:V_e}), with $q_{\text{eq}}=(-15,15)$ and
$\kappa_{e}=R(20^{\circ})^{T}\cdot\Omega\cdot R(20^{\circ})$; the positions,
provided in atomic units, are in the mass-weighted ground-state normal-mode
coordinates with a common scaled mass $m=1$,
\begin{equation}
R(\theta) =
\begin{pmatrix}
\text{cos}(\theta) & -\text{sin}(\theta)\\
\text{sin}(\theta) & \text{cos}(\theta)
\end{pmatrix}
\end{equation}
is the rotation matrix, and $\Omega$ is the diagonal Hessian matrix
corresponding to vibrational wavenumbers $\tilde{\nu}_{1}^{\prime
}=750\,\text{cm}^{-1}$ and $\tilde{\nu}_{2}^{\prime}=2200\,\text{cm}^{-1}$.
The parameters $\Lambda_{t}$ of the Gaussian associated with the Hagedorn
functions~(\ref{eq:phi_k}) were propagated for 20000 steps with a time step
$\Delta t=2\,\text{a.u.}$ (i.e., for a total time $40000\,\text{a.u.}$) using
the exact propagation scheme from Ref.~\onlinecite{Barbiero_Vanicek:2026}. The
numerically exact quantum propagation of the initial states was performed
using the second-order Fourier split-operator
algorithm~\cite{Feit_Steiger:1982_v2} on a position grid ranging from
$q_{\text{eq},j}-128$ to $q_{\text{eq},j}+128$ in each dimension $j=1,2$, with
a total of $256\times256$ equidistant grid points. The computed RR excitation
profiles were broadened using a Gaussian with a half-width at half-maximum of
$50\,\text{cm}^{-1}$, while the RR spectra were constructed from the
excitation profiles assuming Gaussian peaks with a half-width at half-maximum
of $35\,\text{cm}^{-1}$.

\subsection{Anthracene}

To compare with previously reported results and to validate our method, we
used the same density functional theory (DFT) method (at $\omega
$B97X/def2-TZVP level of
theory) as in
Ref.~\onlinecite{DeSouza_Izsak:2019} to construct the global harmonic
surfaces. Solvent effects were included by means of a polarizable continuum
model. Geometry optimizations and frequency calculations were performed using
the Gaussian 16 package~\cite{Frisch_Fox:2016} for the ground [$\text{S}_{0}$
($^{1}\text{A}_{\text{g}}$, point group $D_{2h}$)] and the third excited
[$\text{S}_{3}$ ($^{1}\text{B}_{2\text{u}}$, point group $D_{2h}$)] electronic
states. An empirical scaling factor of 0.95 was applied to the ground-state
vibrational wavenumbers to account for the systematic error of
DFT.~\cite{DeSouza_Izsak:2019} The excited-state calculations were carried out
using standard linear-response time-dependent DFT. The optimized structures of
the two states and the unscaled wavenumbers of the vibrational modes are
listed in the Supporting Information.

The Gaussian wavepacket was propagated for 1000 steps with a time step of
$8\,\text{a.u.}$ (i.e., for a total time of $8000\,\text{a.u.}$) using the
exact propagation scheme from Ref.~\onlinecite{Barbiero_Vanicek:2026}. Based
on the available experimental absorption spectrum,~\cite{Efremov_Gooijer:2006}
the computed RR excitation profiles were broadened using a Voigt function,
i.e. a Gaussian with a half-width at half-maximum of $375\,\text{cm}^{-1}$
multiplied with a Lorentzian with a half-width at half-maximum of
$95\,\text{cm}^{-1}$, and shifted by $-220\,\text{cm}^{-1}$ to account for the
error in the ab initio electronic structure estimate of the vertical
excitation energies (see Fig. S2 of the Supporting Information). In the
simulated RR spectra, individual transitions were convoluted with Gaussian
functions of half-width at half-maximum $12\,\text{cm}^{-1}$ to match the
experimental linewidths.

\section*{Supporting Information}

See the supporting information for (i)~optimized geometries, harmonic
wavenumbers, and absorption spectrum of anthracene, (ii)~analysis of
solvent effects on the resonance Raman spectrum of anthracene,
and (iii)~simplified expressions for specific overlaps of Hagedorn functions.

\section*{Acknowledgments}

This research was supported by the Swiss National Science Foundation (Grant
No. 10005187).

\section*{Data availability}

The data that support the findings of this study are openly available in
Zenodo at https://doi.org/10.5281/zenodo.21885391.

\bibliographystyle{aipnum4-2}
\bibliography{RR_from_HWD}

\end{document}


\title{Supporting Information for ``Resonance Raman spectroscopy from ab initio Hagedorn wavepacket dynamics''}
\author{Davide Barbiero}
\author{L\'{e}a Zupan}
\author{Ji\v{r}\'i J. L. Van\'i\v{c}ek}
\email{jiri.vanicek@epfl.ch}
\affiliation{Laboratory of Theoretical Physical Chemistry, Institut des Sciences et
Ing\'enierie Chimiques, Ecole Polytechnique F\'ed\'erale de Lausanne (EPFL),
CH-1015, Lausanne, Switzerland}
\date{\today}

\begin{abstract}
This document, which provides supplementary information to the main text, contains (i)~optimized geometries, harmonic wavenumbers, and absorption spectrum of anthracene, (ii)~analysis of solvent effects on the resonance Raman spectrum of anthracene, and (iii)~simplified expressions for specific overlaps of Hagedorn functions.
\end{abstract}

\maketitle

\section{Resonance Raman spectrum of anthracene}

\begin{figure} [h]
\includegraphics{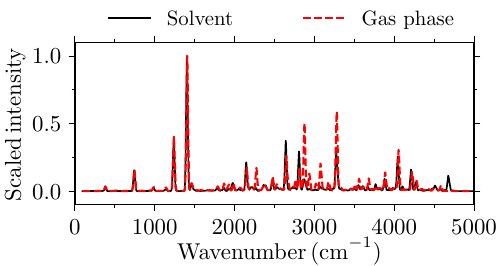}
\caption{Resonance Raman spectrum of anthracene for $\tilde{\nu}_{I}=43668\,\text{cm}^{-1}$. Comparison between spectra computed using Hagedorn wavepacket dynamics at 0 K with and without the inclusion of solvent effects.}
\label{fig:ANT_solv}%
\end{figure}

\begin{figure} [h]
\includegraphics{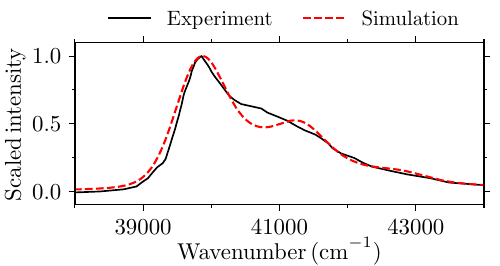}
\caption{Absorption spectrum of anthracene. The adiabatic harmonic spectrum (computed with Gaussian wavepacket dynamics at 0 K) is compared to the experimental spectrum (measured at 233 K).~\cite{Efremov_Gooijer:2006} The computed spectrum was shifted by $-220\,\text{cm}^{-1}$ and broadened using a Voigt function, i.e. a Gaussian with a half-width at half-maximum of $375\,\text{cm}^{-1}$ multiplied with a Lorentzian with a half-width at half-maximum of
$95\,\text{cm}^{-1}$, to maximize the overlap with the experimental spectrum. The same shift and broadening were then used in the main text to evaluate resonance Raman profiles.}
\label{fig:ANT_abs}%
\end{figure}

Equilibrium nuclear geometries in the ground [$\text{S}_{0}\,(^{1}\text{A}_{g}$, point group $D_{2h})$] and third excited [$\text{S}_{3}\,(^{1}\text{B}_{2u}$, point group $D_{2h})$] electronic states of anthracene were calculated using the $\omega$B97X functional and the def2-TZVP basis set. The optimized structures in Cartesian coordinates are provided in Tables~\ref{tab:ground_anth} and~\ref{tab:excited_anth}. The molecule has been placed in the $yz$ plane with the long molecular axis corresponding to the $y$ axis. Table~\ref{tab:freq_anth} presents the harmonic vibrational wavenumbers in the two electronic states. 
Figure~\ref{fig:ANT_solv} shows that solvent appears to have a similar effect on the computed resonance Raman spectrum of anthracene as does the temperature, i.e., the attenuation of overtone and combination-band intensities. Finally, the adiabatic harmonic absorption spectrum is compared with the experimental one in Fig.~\ref{fig:ANT_abs}. 

\begin{table}
    \centering
    \caption{Ground-state equilibrium geometry (in $\text{\r{A}}$) of anthracene at the $\omega$B97X/def2-TZVP level of theory.}
    \label{tab:ground_anth}
    \begin{tabular}{cddd}
         \hline \hline
         &  \multicolumn{1}{c}{$x$} & \multicolumn{1}{c}{$y$} & \multicolumn{1}{c}{$z$}\\
         \hline 
C & 0.0000 & 2.4700 & 1.4010 \\
C & 0.0000 & 3.6383 & 0.7135 \\
C & 0.0000 & 3.6383 & -0.7135 \\
C & 0.0000 & 2.4700 & -1.4010 \\
C & 0.0000 & 1.2144 & -0.7142 \\
C & 0.0000 & 1.2144 & 0.7142 \\
C & 0.0000 & 0.0000 & -1.3950 \\
H & 0.0000 & 4.5833 & -1.2438 \\                  
H & 0.0000 & 2.4658 & -2.4858 \\
H & 0.0000 & 4.5833 & 1.2438 \\
C & 0.0000 & 0.0000 & 1.3950 \\
C & 0.0000 & -1.2144 & 0.7142 \\
H & 0.0000 & 2.4658 & 2.4858 \\
C & 0.0000 & -1.2144 & -0.7142 \\
H & 0.0000 & 0.0000 & -2.4808 \\
C & 0.0000 & -2.4700 & -1.4010 \\
H & 0.0000 & 0.0000 & 2.4808 \\
C & 0.0000 & -2.4700 & 1.4010 \\
C & 0.0000 & -3.6383 & 0.7135 \\                  
C & 0.0000 & -3.6383 & -0.7135 \\
H & 0.0000 & -4.5833 & -1.2438 \\
H & 0.0000 & -2.4658 & -2.4858 \\
H & 0.0000 & -4.5833 & 1.2438 \\
H & 0.0000 & -2.4658 & 2.4858 \\
         \hline \hline
    \end{tabular}
\end{table}
\begin{table}[]
    \centering
        \caption{Excited-state equilibrium geometry (in $\text{\r{A}}$) of anthracene at the\\ TD-$\omega$B97X/def2-TZVP level of theory.}
    \label{tab:excited_anth}
    \begin{tabular}{cddd}
        \hline \hline
         &  \multicolumn{1}{c}{$x$} & \multicolumn{1}{c}{$y$} & \multicolumn{1}{c}{$z$}\\
         \hline 
C & 0.0000 &  2.4716 &  1.4061 \\
C & 0.0000 &  3.6587 &  0.7133 \\
C & 0.0000 &  3.6587 & -0.7133 \\
C & 0.0000 &  2.4716 & -1.4061 \\
C & 0.0000 &  1.2269 & -0.7315 \\
C & 0.0000 &  1.2269 &  0.7315 \\
C & 0.0000 &  0.0000 & -1.4101 \\
H & 0.0000 &  4.6011 & -1.2487 \\
H & 0.0000 &  2.4733 & -2.4911 \\
H & 0.0000 &  4.6011 &  1.2487 \\
C & 0.0000 &  0.0000 &  1.4101 \\
C & 0.0000 & -1.2269 &  0.7315 \\
H & 0.0000 &  2.4733 &  2.4911 \\
C & 0.0000 & -1.2269 & -0.7315 \\
H & 0.0000 &  0.0000 & -2.4956 \\
C & 0.0000 & -2.4716 & -1.4061 \\
H & 0.0000 &  0.0000 &  2.4956 \\
C & 0.0000 & -2.4716 &  1.4061 \\
C & 0.0000 & -3.6587 &  0.7133 \\
C & 0.0000 & -3.6587 & -0.7133 \\
H & 0.0000 & -4.6011 & -1.2487 \\
H & 0.0000 & -2.4733 & -2.4911 \\
H & 0.0000 & -4.6011 &  1.2487 \\
H & 0.0000 & -2.4733 &  2.4911 \\
         \hline \hline
    \end{tabular}
\end{table}

\begin{table}[]
    \centering
     \caption{Computed vibrational wavenumbers (in $\text{cm}^{-1}$) in $\text{S}_{0}$ ($\tilde{\nu}_{j}^{\prime\prime}$) and $\text{S}_{3}$ ($\tilde{\nu}_{j}^{\prime}$) electronic states of anthracene at the $\omega$B97X/def2-TZVP level of theory.}
    \label{tab:freq_anth}
    \begin{tabular}{ccdd|ccdd|ccdd}
        \hline \hline
         Mode $j$ & Symm. & \multicolumn{1}{c}{$\tilde{\nu}_{j}^{\prime\prime}$} & \multicolumn{1}{c}{$\tilde{\nu}_{j}^{\prime}$} & Mode $j$ & Symm. & \multicolumn{1}{c}{$\tilde{\nu}_{j}^{\prime\prime}$} & \multicolumn{1}{c}{$\tilde{\nu}_{j}^{\prime}$} & Mode $j$ & Symm. & \multicolumn{1}{c}{$\tilde{\nu}_{j}^{\prime\prime}$} & \multicolumn{1}{c}{$\tilde{\nu}_{j}^{\prime}$} \\
        \hline 
        1 & a$_{g}$ & 3231 & 3229 & 23 & b$_{2u}$ & 1356 & 1223 & 45 & b$_{2u}$ & 832 & 806 \\
        2 & b$_{2u}$ & 3231 & 3228 & 24 & a$_{g}$ & 1303 & 1300 & 46 & b$_{2g}$ & 810 & 736 \\
        3 & b$_{1u}$ & 3219 & 3215 & 25 & b$_{3g}$ & 1308 & 1276 & 47 & b$_{1g}$ & 789 & 726 \\
        4 & b$_{3g}$ & 3219 & 3216 & 26 & b$_{1u}$ & 1302 & 969 & 48 & a$_{g}$ & 781 & 750 \\
        5 & a$_{g}$ & 3205 & 3204 & 27 & b$_{3g}$ & 1216 & 1217 & 49 & a$_{u}$ & 785 & 695 \\
        6 & b$_{2u}$ & 3204 & 3201 & 28 & b$_{2u}$ & 1193 & 1417 & 50 & b$_{3u}$ & 757 & 717 \\
        7 & b$_{1u}$ & 3201 & 3199 & 29 & a$_{g}$ & 1189 & 1183 & 51 & b$_{1u}$ & 665 & 541 \\
        8 & b$_{3g}$ & 3201 & 3198 & 30 & b$_{1u}$ & 1182 & 1201 & 52 & a$_{g}$ & 643 & 624 \\
        9 & a$_{g}$ & 3197 & 3199 & 31 & b$_{3g}$ & 1141 & 1126 & 53 & b$_{2u}$ & 617 & 610 \\
        10 & b$_{1u}$ & 3195 & 3197 & 32 & b$_{2u}$ & 1130 & 1170 & 54 & b$_{2g}$ & 604 & 490 \\
        11 & b$_{3g}$ & 1731 & 1874 & 33 & a$_{g}$ & 1038 & 1028 & 55 & b$_{3g}$ & 540 & 527 \\
        12 & b$_{1u}$ & 1722 & 1329 & 34 & b$_{2u}$ & 1029 & 1015 & 56 & a$_{u}$ & 514 & 442 \\
        13 & b$_{3g}$ & 1669 & 1566 & 35 & b$_{2g}$ & 1034 & 950 & 57 & b$_{1g}$ & 497 & 443 \\
        14 & a$_{g}$ & 1651 & 1605 & 36 & a$_{u}$ & 1034 & 946 & 58 & b$_{3u}$ & 490 & 432 \\
        15 & b$_{2u}$ & 1622 & 1573 & 37 & b$_{3u}$ & 1019 & 961 & 59 & a$_{g}$ & 401 & 399 \\
        16 & a$_{g}$ & 1537 & 1534 & 38 & b$_{1g}$ & 1015 & 955 & 60 & b$_{3g}$ & 403 & 391 \\
        17 & b$_{1u}$ & 1509 & 1504 & 39 & b$_{2g}$ & 955 & 889 & 61 & b$_{3u}$ & 393 & 344 \\
        18 & b$_{2u}$ & 1498 & 1494 & 40 & b$_{3g}$ & 935 & 922 & 62 & b$_{2g}$ & 272 & 204 \\
        19 & a$_{g}$ & 1478 & 1449 & 41 & b$_{3u}$ & 938 & 876 & 63 & b$_{1u}$ & 239 & 234 \\
        20 & b$_{2u}$ & 1458 & 1478 & 42 & b$_{1u}$ & 926 & 875 & 64 & b$_{1g}$ & 236 & 220 \\
        21 & b$_{3g}$ & 1425 & 1411 & 43 & a$_{u}$ & 895 & 834 & 65 & a$_{u}$ & 123 & 86 \\
        22 & b$_{1u}$ & 1362 & 1382 & 44 & b$_{2g}$ & 872 & 810 & 66 & b$_{3u}$ & 90 & 85 \\
        \hline \hline
    \end{tabular}
\end{table}


\section{Overlap of Hagedorn functions}

The cross-correlation functions needed for evaluating the resonance Raman profiles require evaluating the overlap matrix
\begin{equation}
    M_{JK^{\prime}}:=\langle J(\Lambda)|K^{\prime}(\Lambda^{\prime})\rangle
\end{equation}
of Hagedorn functions with different Gaussian centers, where we have introduced the notation $J(\Lambda):=\varphi_{J}[\Lambda_{0}]\equiv\psi_{f}(0)$ and $K^{\prime}(\Lambda^{\prime}):=\varphi_{K}[\Lambda_{t}]\equiv\psi_{i}(t)$.

The recursive expressions from Ref.~\citenum{Vanicek_Zhang:2025_v2} are valid for any excitation shell in the stationary, $J(\Lambda)$, and evolved, $K^{\prime}(\Lambda^{\prime})$, Hagedorn functions, as well as for arbitrary parameters $\Lambda$ and $\Lambda^{\prime}$ not related by time evolution.
Here, we describe how the general expressions simplify if one is interested in low excitations (i.e., fundamentals, first overtones, and first combination bands) in the zero-temperature resonance Raman spectrum. 

\subsection{Zero-temperature approximation}

Within the zero-temperature approximation, one needs to compute the ``vector'' of overlaps
\begin{equation}
    d_{J}:=M_{J0^{\prime}}=\langle J(\Lambda)|0^{\prime}(\Lambda^{\prime})\rangle
\end{equation}
of the propagated Gaussian $0^{\prime}(\Lambda^{\prime})$ with Hagedorn functions $J(\Lambda)$ associated with the initial Gaussian $0(\Lambda)$.
All overlaps $d_{J}$ can be obtained from $d_{0}$ [Eq.~(80) of Ref.~\citenum{Vanicek_Zhang:2025_v2}] using the recurrence relation~\cite{Vanicek_Zhang:2025_v2}
\begin{equation}
    \sqrt{J_{j}+1}d_{J+\langle j\rangle}=u_{j}d_{J}-\sum_{k=1}^{D}G_{jk}\sqrt{J_{k}}d_{J-\langle k\rangle},\label{eq:d_J}
\end{equation}
This result is a special case $d_{J}:=M_{J0^{\prime}}$ of a general result for the overlap $M_{JK^{\prime}}$ proven in Proposition 10 of Ref.~\citenum{Vanicek_Zhang:2025_v2}. 
The $D\times D$ matrix $G$ and the $D$-dimensional vector $u$ are defined in Eqs.~(102) and~(104) of Ref.~\citenum{Vanicek_Zhang:2025_v2}. These variables are independent of multi-index $J$ and depend only on the guiding Gaussians.

\subsubsection{Fundamentals (first shell)}

Setting $J:=0$ in the general expression~(\ref{eq:d_J}) for $d_{J+\langle j\rangle}$ yields~\cite{Vanicek_Zhang:2025_v2}
\begin{equation}
    d_{\langle j \rangle}=u_{j}d_{0},\quad j=1,\dots,D.
\end{equation}
In vector form, the solution can be written as $d_{1}=ud_{0}$, where $d_{1}:=d_{|J|=1}$ is the $D$-vector of overlaps.

\subsubsection{First overtones and combination bands (second shell)}

Setting $J:=\langle l\rangle$ in the general expression~(\ref{eq:d_J}) for $d_{J+\langle j\rangle}$ yields~\cite{Vanicek_Zhang:2025_v2}
\begin{equation}
    \sqrt{\delta_{jl}+1}d_{\langle j\rangle+\langle l\rangle}=(u\otimes d_{1}^{T}-Gd_{0})_{jl}=(u\otimes u^{T}-G)_{jl}d_{0}
\end{equation}
for the overlap vector $d_{2}:=d_{|J|=2}$.

\bibliographystyle{aipnum4-2}
\bibliography{RR_from_HWD}